\documentclass[conference]{IEEEtran}
\IEEEoverridecommandlockouts
\usepackage[colorlinks=true, linkcolor=black, urlcolor=black, citecolor=black]{hyperref}

\usepackage{float}
\usepackage{lipsum}
\usepackage{cite}
\usepackage{amsmath,amssymb,amsfonts}
\usepackage{algorithmic}
\usepackage{graphicx}
\usepackage{textcomp}
\usepackage{tabularx}
\usepackage{xcolor}

\def\BibTeX{{\rm B\kern-.05em{\sc i\kern-.025em b}\kern-.08em
    T\kern-.1667em\lower.7ex\hbox{E}\kern-.125emX}}
\begin{document}
\title{A Two-Stage Band-Split Mamba-2 Network For Music Separation \\
}
\author{
    Jinglin Bai\textsuperscript{1,2,$\dagger$}, Yuan Fang\textsuperscript{1,2,$\dagger$}\thanks{$\dagger$ Work done during internship at SenseTime.}, Jiajie Wang\textsuperscript{2}, Xueliang Zhang\textsuperscript{1}\\
    \textsuperscript{1}College of Computer Science, Inner Mongolia University, Hohhot, China \\
    \textsuperscript{2}SenseAuto Intelligent Connection, SenseTime, Beijing, China\\
    \{bjlin, 32209021\}@mail.imu.edu.cn, wangjiajie1@senseauto.com, cszxl@imu.edu.cn
}

\maketitle

\begin{abstract}

Music source separation (MSS) aims to separate mixed music into its distinct tracks, such as vocals, bass, drums, and more. MSS is considered to be a challenging audio separation task due to the complexity of music signals.  Although the RNN and Transformer architecture are not perfect, they are commonly used to model the music sequence for MSS. Recently, Mamba-2 has already demonstrated high efficiency in various sequential modeling tasks, but its superiority has not been investigated in MSS. This paper applies Mamba-2 with a two-stage strategy, which introduces residual mapping based on the mask method, effectively compensating for the details absent in the mask and  further  improving  separation performance. Experiments confirm the superiority of bidirectional Mamba-2 and the effectiveness of the two-stage network in MSS. The source code is publicly accessible at
\texttt{\url{https://github.com/baijinglin/TS-BSmamba2}}.

\end{abstract}


\section{Introduction} 
Music source separation (MSS) separates distinct tracks from mixed music. The MSS system has wide applicability, such as  music information retrieval
 [\citenum{mesaros2010automatic}, \citenum{itoyama2011query}], music remixing \cite{gillet2005extraction, pons2016remixing, woodruff2006remixing}, and music education [\citenum{cano2014pitch}, \citenum{dittmar2012music}]. Many deep learning methods have been proposed to tackle the MSS problem, such as frequency-domain systems \cite{uhlich2017improving,chandna2017monoaural,hennequin2020spleeter,takahashi2018mmdenselstm,li2021sams,kong2021decoupling}, time-domain systems\cite{defossez2019music, samuel2020meta, stoller2018wave}, 
and the cross-domain systems \cite{rouard2023hybrid, jung2021kuielab, defossez2021hybrid, venkatesh2024real}.

  Recently, band-split RNN (BSRNN) \cite{luo2023music} has achieved landmark results, which splits the complex-valued mixture spectrogram into sub-band spectrograms and performs sequential sequence-level and band-level modeling using residual bidirectional LSTM (BLSTM) layers. As a continuation, the Single-input-multi-output (SIMO) stereo BSRNN \cite{luo2024improving} is proposed to predict all tracks with a unified model while considering the correlation of channels. The advancement of BSRNN has also motivated many studies. Some focus on exploring lightweight networks, such as the dual-path TFC-TDF UNet (DTTNet) \cite{chen2024music} and the sparse compression network (SCNet) \cite{tong2024scnet}. Others aim to achieve superior results without considering the computational cost. For example, the Band-Split RoPE Transformer (BS-RoFormer) \cite{lu2024music} combines the Transformer \cite{vaswani2017attention} with Rotary Position Embedding (RoPE) \cite{su2024roformer} to replace the RNN.
  
 
Compared to speech, music exhibits greater complexity and stronger long-term dependencies, which are often processed by the RNN  or Transformer architecture for MSS.  However, RNNs are hard to parallelize and struggle with vanishing or exploding gradients. Transformers require quadratic computation for self-attention, which often results in computational load. Recently, Mamba-2 \cite{mamba2} has been proposed as a promising approach to replace RNN and Transformer architectures, building on the state space model (SSM) \cite{gu2021efficiently} and Mamba-1 \cite{gu2023mamba}. Mamba-2 introduces a novel connection between SSMs and  attention variants called structured state space duality (SSD). SSD is based on block decompositions of semiseparable matrices \cite{vandebril2005bibliography}, utilizing the linear SSM recurrence and quadratic dual form to achieve a new efficient and easily-implementable algorithm for computing SSMs.

  \begin{figure*}[ht]
    \centering
\includegraphics[width=0.96\linewidth, trim={0.00cm 0.5cm 0.00cm 3.0 cm}, clip]{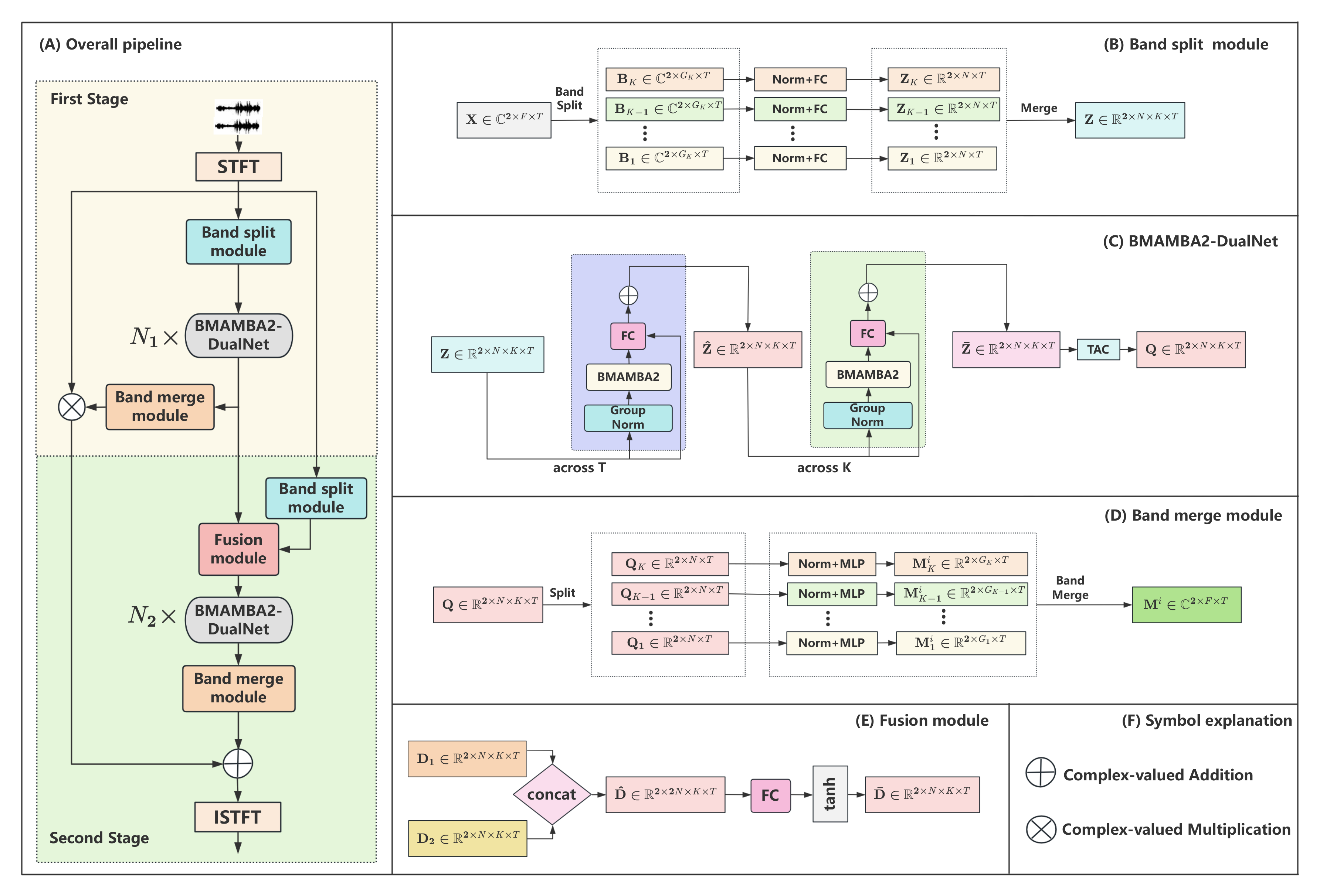}
    \caption{(A) The overall pipeline of TS-BSMAMBA2. (B) The design of the band split module. (C) The BMAMBA2-DualNet is designed based on the BMAMBA2 structure. (D) The design of the band merge module, with each playing a different role in the two stages. (E) The design of the fusion module is to integrate the features from the first and second stages. (F) Symbol explanation of overall pipeline.}
    \label{fig:TS-BSMAMBA2}
\end{figure*}
Current solutions for MSS are primarily single-stage mask [\citenum{luo2023music}, \citenum{luo2024improving}, \citenum{lu2024music}] or mapping [\citenum{chen2024music}, \citenum{tong2024scnet}] methods, with mask methods being the mainstream. In other tasks\cite{9802706,li2022taylorhearnowtaylorunfolding,li2022taylorbeamformer}, the multi-stage method combining mask and mapping has been proven to outperform the single-stage mask method. Inspired by this, we apply a two-stage method to network design. The first stage estimates complex masks for different tracks to learn coarse features. Building on the first stage, the second estimates residual mappings to capture fine-grained features. To the best of our knowledge, we are the first to introduce the two-stage method to MSS. This paper proposes a novel two-stage band-split Mamba-2 separation network (TS-BSMAMBA2).  The contributions of our paper are as follows:
\begin{enumerate}
    \item We are the first to utilize Mamba-2 for MSS.
    \item The effectiveness of the two-stage method in MSS has been validated for the first time by our experiments.  
    \item Our proposed method achieves superior performance with reduced computation and fewer parameters while delivering good results in lightweight models.
\end{enumerate}

\section{TS-BSMAMBA2}


As shown in Fig.~\ref{fig:TS-BSMAMBA2} (A), a dual-channel input is first processed through Short-Time Fourier Transform (STFT) to obtain $\mathbf{X} \in \mathbb{C}^{2 \times F \times T}$, where $F$ and $T$ represent the number of frequency bins and time frames, respectively. TS-BSMAMBA2 consists of two stages. In the first stage, complex T-F masks are applied to predict different music tracks. In the second stage, residual mappings are predicted through the network, which is then added to the first stage outputs. Finally, Inverse Short-Time Fourier Transform (ISTFT) is applied to restore the dual-channel audio for four tracks.
\subsection{Band split module}
\label{ssec:subhead}
The band-split scheme [\citenum{luo2023music}, \citenum{luo2024improving}, \citenum{tong2024scnet}, \citenum{lu2024music}] has effectively improved model performance in MSS tasks.
As illustrated in Fig.~\ref{fig:TS-BSMAMBA2} (B), the complex-valued stereo input mixture spectrogram \( \mathbf{X} \in \mathbb{C}^{2 \times F \times T} \) is split into \( K \) subband spectrograms \( \mathbf{X}_k \in \mathbb{C}^{2 \times G_k \times T}, k = 1, \ldots, K \), each with varying bandwidth \( G_k \). The real and imaginary components of the subband spectrogram for each channel are then concatenated along the frequency axis. After applying a GroupNorm layer \cite{wu2018group} and a fully connected (FC) layer, the features for each subband, denoted as \( \{Z_k\}_{k=1}^{K} \in \mathbb{C}^{2 \times N \times T} \), are obtained, where \( N \) represents the feature dimension. Finally, these features are stacked into a 4-dimensional tensor \( Z \in \mathbb{C}^{2 \times N \times K \times T} \).

\subsection{BMAMBA2-DualNet}
\label{ssec:subhead}
 To  model music sequences more efficiently, two residual bidirectional Mamba-2 (BMAMBA2) layers are sequentially applied across the temporal dimension \( T \) and the band dimension \( K \) to generate the transformed tensor \( \hat{Z} \in \mathbb{R}^{2 \times N \times K \times T} \), as shown in Fig.~\ref{fig:TS-BSMAMBA2} (C). A transform-average-concatenate (TAC) module \cite{luo2020end} is then applied to model the inter-channel dependencies of stereo signals, yielding the tensor \( \mathbf{Q} \in \mathbb{R}^{2 \times N \times K \times T} \). Each residual BMAMBA2 layer consists of a GroupNorm layer, a BMAMBA2 block, and a FC layer. 
 
\begin{figure}[ht]
    \centering
    \includegraphics[width=1.00\linewidth, trim={0.50cm 6.4cm 0.00cm 1.00cm}, clip]{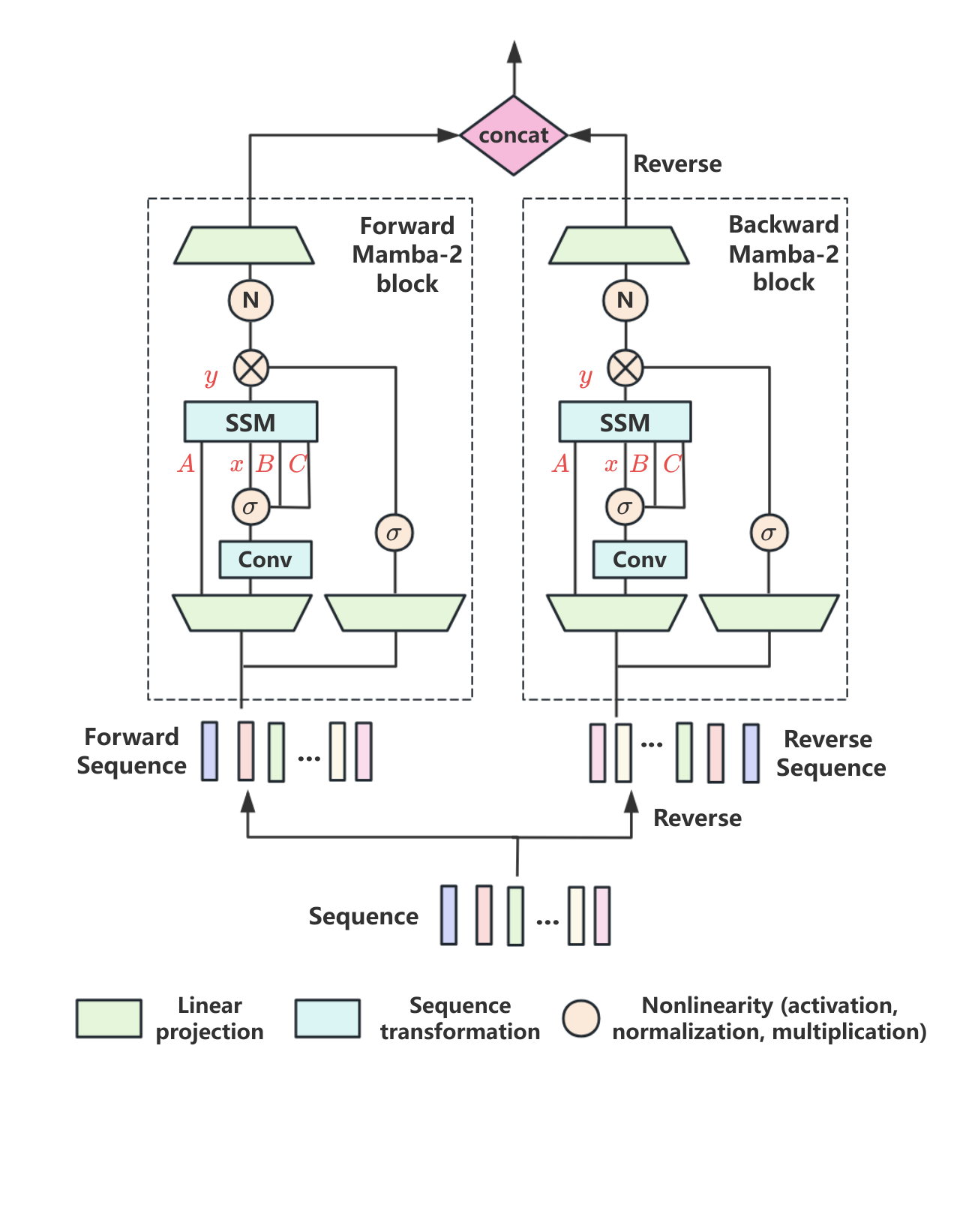}
    \caption{Structure of the BMAMBA2 block}
    \label{fig:BSMAMBA2}
\end{figure}

As  shown in Fig.~\ref{fig:BSMAMBA2}, the BMAMBA2 block contains a forward Mamba2 block and a backward Mamba2 block that model the sequence in forward and backward directions, respectively, and are then concatenated along the  N dimension. The forward and backward Mamba-2 blocks are designed as in \cite{mamba2}.  Mamba-2 is designed based on the SSD framework, whose core layer is a refinement of Mamba's selective SSM. Inspired by the continuous system, SSM maps an input function \( \mathbf{x}(t) \in \mathbb{R} \) to an output \( \mathbf{y}(t) \in \mathbb{R} \) through a higher-dimensional latent state \( \mathbf{h}(t) \in \mathbb{R}^{{H}} \) over all continuous time steps \( T \), where \( {H} \) denotes the number of feature channels. For discrete-time signals in digital devices, the discrete selective SSM updates at each time step $t$ with discretized matrices $\overline{\mathbf{A}}$ and $\overline{\mathbf{B}}$, as shown in (1) and (2).
 \begin{align}
    \mathbf{h}_t &= \overline{\mathbf{A}}\mathbf{h}_{t-1} + \overline{\mathbf{B}}\mathbf{x}_t, \quad \mathbf{y}_t = \mathbf{C}\mathbf{h}_t \tag{1}.
\end{align}

\begin{align}
    \overline{\mathbf{A}} &= \exp(\Delta \mathbf{A}), \quad \overline{\mathbf{B}} = (\Delta \mathbf{A})^{-1}(\exp(\Delta \mathbf{A}) - \mathbf{I})  \Delta \mathbf{B} \tag{2}.
\end{align}

Here, $\mathbf{A} \in \mathbb{R}^{H \times H}$ denotes the evolution parameter, $\mathbf{B} \in \mathbb{R}^{H \times 1}$  and $\mathbf{C} \in \mathbb{R}^{H \times 1}$ denote the projection parameters. $\overline{\mathbf{A}}$ and $\overline{\mathbf{B}}$ are approximated by zero-order hold discretization formulas. A learnable parameter $\Delta$ balances how much to focus on or ignore the current state and input. 

\subsection{Band merge module} \label{ssec:subhead}
As shown in Fig.~\ref{fig:TS-BSMAMBA2} (D), for target source \(i \in [1, 4]\), the subband feature \( \mathbf{Q}_k \in \mathbb{R}^{2 \times N \times T} \) from  BMAMBA2-DualNet is passed to the \(i\)-th layer normalization followed by the \(i\)-th multilayer perceptron (MLP) with gated linear unit (GLU) activation \cite{dauphin2017language} to generate the real and imaginary components of the \(i\)-th complex-valued time-frequency (T-F) tensor \( \mathbf{M}_k^i \) at band \(k\). 
As shown in Fig.~\ref{fig:TS-BSMAMBA2} (A), the band merge module in the first stage is  applied to estimate complex T-F masks and then multiplied with the mixture \( \mathbf{X} \) in the complex domain. In the second stage,  the band merge module generates residual mappings, which are then added to the first stage output.

\subsection{Fusion Module}
As shown in Fig.~\ref{fig:TS-BSMAMBA2} (E), $D_1$ $\in \mathbb{R}^{2 \times N \times K \times T}$ is the output from Dual-path Module of the first stage, and $D_2$ $\in \mathbb{R}^{2 \times N \times K \times T}$ is the output from band split module of the second stage. We concatenate them along the $N$ dimension to form $\hat{D}$ $\in \mathbb{R}^{2 \times 2N \times K \times T}$. After passing through a FC layer followed by a $\tanh$ activation function, which reduces the $2N$ dimension back to $N$, $\bar{D} \in \mathbb{R}^{2 \times N \times K \times T}$ is obtained.

\section{Experimental Configuration}
\label{sec:print}

\subsection{Data preparation and simulation}
\label{ssec:subhead}
We evaluate our proposed method on MUSDB18-HQ \cite{rafii2019musdb18}, which contains \textit{vocals}, \textit{bass}, \textit{drums}, and \textit{other} tracks. Following the same pipeline as \cite{luo2023music}, we utilize an energy-based unsupervised source activity detector (SAD) to preprocess the training data by excluding silent regions. In the training set of the dataset, the first 85\% of segments from all tracks of each song are used for model training, while the last 15\% are reserved for model validation. We apply batch-level on-the-fly data simulation \cite{uhlich2017improving} by randomly mixing sound tracks from different songs, consistent with the training conditions in \cite{luo2024improving}.


\begin{table*}[t]
\centering
\caption{Model comparison on MUSDB18-HQ  dataset. cSDR and uSDR scores are reported on decibel scale.}
\resizebox{\textwidth}{!}{
\normalsize
\begin{tabular*}{\textwidth}{p{5cm}|cc|cc|cc|cc|cc} 
\hline
 \rule{0pt}{2.5ex} Model & \multicolumn{2}{c|}{Vocals} & \multicolumn{2}{c|}{Bass} & \multicolumn{2}{c|}{Drums} & \multicolumn{2}{c|}{Other} & \multicolumn{2}{c}{Average} \\ \cline{2-11} 
                       & cSDR & uSDR & cSDR & uSDR & cSDR & uSDR & cSDR & uSDR & cSDR & uSDR \\ \hline
KUIELab-MDX-Net \cite{jung2021kuielab} & 8.97 & -- & 7.83 & -- & 7.20 & -- & 5.90 & -- & 7.47 & -- \\ 
Hybrid Demucs \cite{defossez2021hybrid}   & 8.13 & -- & 8.76 & -- & 8.24 & -- & 5.59 & -- & 7.68 & -- \\ 
HT Demucs \cite{rouard2023hybrid}      & 7.93 & -- & 8.48 & -- & 7.94 & -- & 5.72 & -- & 7.52 & -- \\
TFC-TDF-UNet v3 \cite{kim2023sound} &9.59 & --  &8.45  &--  &8.44  & --   &6.86  &--  &8.34  & --\\ 
DTTNet \cite{chen2024music}      &10.12  & -- &7.45  &-- &7.74  & --   &6.92  &--  &8.06  & --\\ 
SCNet \cite{tong2024scnet}    &9.89  & -- &8.82  & -- &\textbf{10.51}  &--  &6.76  &--  &9.00 & -- \\ 
BSRNN \cite{luo2023music} & 10.01 & 10.07 & 7.22 & 6.80 & 9.01 & 8.92 & 6.70 & 6.01 & 8.24 & 7.94 \\ 
SIMO stereo BSRNN \cite{luo2024improving} & 9.73 & 10.27 & 7.80 & \textbf{7.61} & 10.06 & 9.83 & 6.56 & 6.50 & 8.54 & 8.55 \\ \hline

 \rule{0pt}{2.5ex}TS-BSMAMBA2 (FS) &9.77 & 10.11  & 7.56 & 6.68 &9.49  &9.48  &7.53  & 6.28  &8.59  & 8.14  \\ 
TS-BSMAMBA2 (SS)  &\textbf{10.57} & \textbf{10.60} & \textbf{8.88} & 7.47 & 10.34&\textbf{10.07} & \textbf{8.45} & \textbf{6.69} & \textbf{9.56} & \textbf{8.71}   \\ \hline

 \rule{0pt}{2.5ex}L-TS-BSMAMBA2 (FS) & 9.46 & 9.43 &  6.98 & 6.24 & 9.14 &  9.04& 7.16 &6.00 & 8.19 & 7.68  \\ 
L-TS-BSMAMBA2 (SS)   &10.03 & 9.90  & 8.25& 6.91 & 9.79 &  9.56 & 8.04 & 6.39 &9.03  & 8.19   \\ \hline
\end{tabular*} 
}
\label{tab:table1}
\end{table*}

\subsection{Model configuration}
\label{ssec:subhead}
The proposed method uses the same band-split scheme as \cite{luo2024improving}, consisting of 57 sub-bands. The 2048-point STFT with a 512-point hop size and a Hanning window is applied. The feature dimension \(N\) is set to 128 for each channel. The first stage contains  8-layer MAMBA2-DualNet, while the second stage contains  4-layer MAMBA2-DualNet. In BMAMBA2, the model dimension (\(d_{\text{model}}\)) and state dimension (\(d_{\text{state}}\)) are both set to 128, the convolutional dimension (\(d_{\text{conv}}\)) and expansion factor are set to 4, and the head dimension is set to 64. The TAC hidden dimension is \(3N = 384\), and the hidden dimension of each source’s band merge estimation MLP module is set to \(N = 128\).
The training objective contains the outputs from two stages, where the total loss is the sum of the losses from each stage:
\begin{align}
\mathcal{L}_{\text{total}} = \mathcal{L}_{\text{stage1}} + \mathcal{L}_{\text{stage2}} \tag{3}.
\end{align}
For both stages, the loss is computed across all four sources as the sum of the frequency-domain and time-domain mean absolute error (MAE). Each stage loss \(\mathcal{L}_{\text{stage}}\) is defined as:
\begin{align}
\mathcal{L}_{\text{stage}} = &\sum_{i=1}^{4} |\mathcal{R}(\mathbf{S}^i) - \mathcal{R}(\bar{\mathbf{S}}^i)|_1 + |\mathcal{I}(\mathbf{S}^i) - \mathcal{I}(\bar{\mathbf{S}}^i)|_1 \notag \\
&+ |\text{ISTFT}(\mathbf{S}^i) - \text{ISTFT}(\bar{\mathbf{S}}^i)|_1 \tag{4},
\end{align}
where \(\bar{\mathbf{S}}^i\) and \(\mathbf{S}^i\) are the complex-valued spectrograms of the estimated and clean targets for the \(i\)-th source, \(\mathcal{R}\) and \(\mathcal{I}\) represent the real and imaginary parts, respectively, and ISTFT denotes the inverse short-time Fourier transform. All models are trained for 100 epochs with the Adam optimizer \cite{kingma2014adam}, with an initial learning rate of 1e-3. Each epoch contains 100,000 randomly generated training samples with a batch size of 16, and we use 8 Nvidia V100 GPUs for training. If the model fails to converge for two consecutive epochs, the learning rate is decayed by 0.8. Gradient clipping with a maximum gradient norm of 5 is applied.

\subsection{Evaluation and metrics}
\label{ssec:subhead}

During model inference, the full-length song is segmented into 3 second lengths with a 0.5 second hop size, and each segment is processed independently. We use Source-to-Distortion Ratio (SDR) as an evaluation metric. Chunk-level SDR (cSDR) calculates SDR on 1-second chunks and reports the median value, which is calculated by the standard SDR metric in \textit{bass\_eval} metrics \cite{vincent2006performance}. Utterance-level SDR (uSDR) calculates the SDR for each song and reports its mean value, which is used as the default evaluation metric in the Music Demixing (MDX) Challenge 2021 \cite{mitsufuji2022music}.



\section{Results and Analysis}
\label{sec:print}
To ensure a fair comparison, all models are trained on the MUSDB18-HQ training set and evaluated on its testing set, as shown in Table \ref{tab:table1}. L-TS-BSMAMBA2 represents the lightweight TS-BSMAMBA2, with the key difference being that the first stage uses a 4-layer BMAMBA2-DualNet, and the second stage uses a 2-layer BMAMBA2-DualNet. Table \ref{tab:table2} shows the parameters and Multiply-Accumulate Computations (MACs) of our proposed method compared with recent models, which are calculated for 1s dual-channel music. The MACs of the Mamba-2 block is manually calculated based on the source code.\footnote{\url{https://github.com/state-spaces/mamba/blob/main/mamba_ssm/modules/mamba2.py}}  In Tables \ref{tab:table1} and \ref{tab:table2}, (FS) and (SS) represent the output results of the model's first and second stages, respectively.


 \begin{table}[h]
\centering
\caption{Comparison of recent model parameters and MACs.}
\resizebox{\columnwidth}{!}{
\begin{tabular}{lcc}
\hline
\rule{0pt}{2.5ex}\textbf{Model} & \textbf{Parameters (M)} & \textbf{MACs (G/s)} \\ \hline
\rule{0pt}{2.5ex}BSRNN \cite{luo2023music} & 146.57 & 611.01 \\
SIMO stereo BSRNN \cite{luo2024improving} & 107.73 & 245.58 \\
SCNet \cite{tong2024scnet} & 10.08 & - \\
DTTNet \cite{chen2024music} & 20.00 & 143.23 \\ \hline
\rule{0pt}{2.5ex}TS-BSMAMBA2 (FS) &20.34  &140.61  \\
TS-BSMAMBA2 (SS) & 35.52 & 212.11  \\ \hline
\rule{0pt}{2.5ex}L-TS-BSMAMBA2 (FS) & 15.14 & 71.17 \\
L-TS-BSMAMBA2 (SS) & 27.71 & 107.95 \\ \hline
\end{tabular}
}
\label{tab:table2}
\end{table}
From Tables \ref{tab:table1} and \ref{tab:table2}, it can be observed that TS-BSMAMBA2 achieves higher average cSDR and uSDR across all tracks compared to other baselines, with most tracks showing stronger performance, especially the vocals track. The bass track’s uSDR is slightly below that of SIMO stereo BSRNN, and the drums track’s cSDR is below that of SCNet. Compared to BSRNN and SIMO stereo BSRNN, TS-BSMAMBA2 achieves superior performance with fewer parameters and lower computational complexity, highlighting the efficiency of Mamba-2 over RNN for MSS. L-TS-BSMAMBA2 outperforms other baselines on average  cSDR across all tracks. This reveals the advantages of the lightweight model L-TS-BSMAMBA2.
\begin{figure}[ht]
    \centering
    \includegraphics[width=0.97\linewidth, trim={0.50cm 16.9cm 0.00cm 0.00cm}, clip]{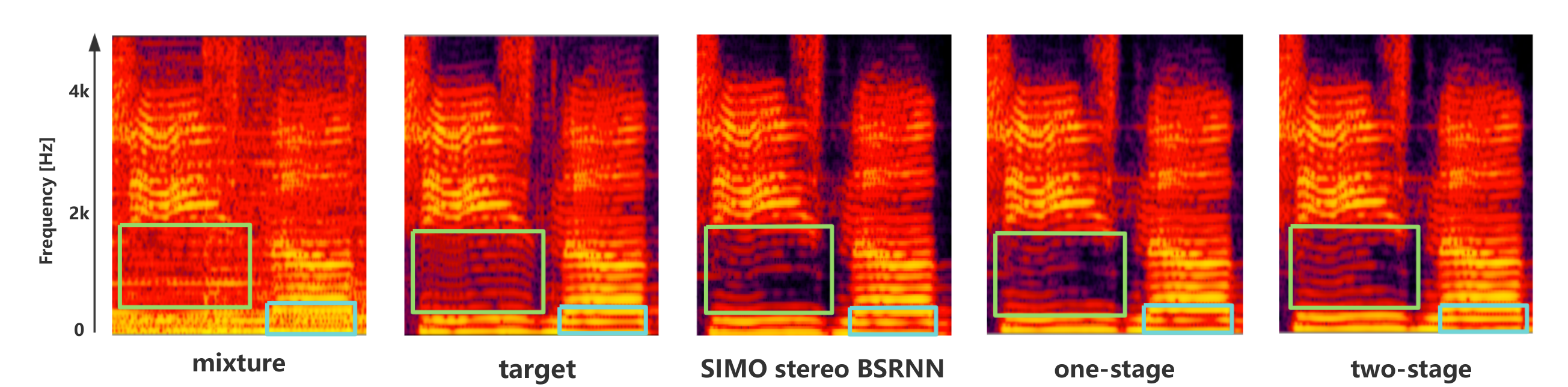}
    \caption{Spectrogram examples of the vocal track.}
    \label{fig:visual}
\end{figure}

TS-BSMAMBA2 (SS) shows performance improvements across all tracks compared to TS-BSMAMBA2 (FS), and similarly, L-TS-BSMAMBA2 (SS) outperforms L-TS-BSMAMBA2 (FS). Fig. \ref{fig:visual} shows the spectrogram examples of SIMO stereo BSRNN and the two-stage outputs from TS-BSMAMBA2. By comparing with SIMO stereo BSRNN and our model first stage output, it is observed that the second stage effectively supplements the missing information from the single-stage mask method by learning residual mappings, demonstrating the effectiveness of the two-stage approach.

\section{conclusion}
This paper introduces a novel two-stage frequency-domain separation network based on the Mamba-2 architecture with the band-split scheme. By utilizing Mamba-2, our proposed approach achieves excellent performance in MSS with high computational efficiency. The effectiveness of the two-stage method in MSS has been demonstrated for the first time through our experiments. Meanwhile, our proposed lightweight method also achieves outstanding cSDR results.


\bibliographystyle{IEEEtran}
\end{document}